%% file: main.tex
\documentclass[sigconf]{acmart}

\usepackage{booktabs}  
\usepackage{multirow}  
\usepackage{makecell}  
\usepackage{graphicx}  
\usepackage{makecell}
\definecolor{tablegreen}{RGB}{173, 216, 230}
\usepackage[table]{xcolor}
\usepackage{pifont}

\usepackage[framemethod=tikz]{mdframed}
\mdfdefinestyle{customstyle}{
  linecolor=gray!100,
  linewidth=3pt,
  innerleftmargin=3pt,
  topline=false,
  rightline=false,
  bottomline=false,
  leftline=true,
  innerrightmargin=3pt,
  innertopmargin=3pt,
  innerbottommargin=3pt,
  backgroundcolor=gray!15
}
\newenvironment{custommdframed}
  {\begin{mdframed}[style=customstyle]}
  {\end{mdframed}}

\usepackage{multirow}
\usepackage[most]{tcolorbox}
\usepackage{pifont}
\usepackage{graphicx}

\usepackage{xspace}
\usepackage{enumitem}
\usepackage{colortbl}

\usepackage{hyperref}
\usepackage{algorithm}
\usepackage{algpseudocode}
\usepackage{array} 
\usepackage{multirow}
\usepackage{xcolor}
\usepackage{threeparttable}
\usepackage[most]{tcolorbox}
\usepackage{listings}
\usepackage{lipsum}
\tcbuselibrary{listingsutf8}
\usepackage{nameref}

\usepackage{caption} 

\usepackage{booktabs}      
\usepackage{graphicx}      
\AtBeginDocument{%
  }

\setcopyright{acmlicensed}
\copyrightyear{2018}
\acmYear{2018}
\acmDOI{XXXXXXX.XXXXXXX}
\acmConference[Conference acronym 'XX]{Make sure to enter the correct
  conference title from your rights confirmation email}{June 03--05,
  2018}{Woodstock, NY}
\acmISBN{978-1-4503-XXXX-X/2018/06}

\newcommand{\eg}{{\emph{e.g.,}}\xspace}
\newcommand{\ie}{{\emph{i.e.,}}\xspace}

\begin{document}



\title{Bridging the Gap between User Intent and LLM: A Requirement Alignment Approach for Code Generation}



\author{Jia Li}
\authornote{Both authors contributed equally to this research.}
\affiliation{%
  \institution{School of Computer Science, \\ Wuhan University}
  \city{Wuhan}
  \country{China}
  }
\email{jia.li@whu.edu.cn}

\author{Ruiqi Bai}
\authornotemark[1]
\affiliation{%
   \institution{School of Computer Science, \\ Wuhan University}
  \city{Wuhan}
  \country{China}
  }
\email{2023302111114@whu.edu.cn}

\author{Yangkang Luo}
\affiliation{%
   \institution{School of Computer Science, \\ Wuhan University}
  \city{Wuhan}
  \country{China}
  }
\email{2024302111342@whu.edu.cn}

\author{Yiran Zhang}
\affiliation{%
   \institution{Nanyang Technological University}
   \country{Singapore}
  }
\email{yiran002@e.ntu.edu.sg}

\author{Wentao Yang}
\affiliation{%
   \institution{School of Computer Science, \\ Wuhan University}
  \city{Wuhan}
  \country{China}
  }
\email{2023302111181@whu.edu.cn}

\author{Zeyu Sun}
\affiliation{%
   \institution{Institute of Software, \\ Chinese Academy of Sciences}
  \city{Beijing}
  \country{China}
  }
\email{zeyu.zys@gmail.com}

\author{Tiankuo Zhao}
\affiliation{%
   \institution{School of Computer Science, \\ Wuhan University}
  \city{Wuhan}
  \country{China}
  }
\email{tiankuoz@whu.edu.cn}

\author{Dongming Jin}
\affiliation{%
   \institution{Key Lab of High Confidence Software Technology, \\ MoE (Peking University)}
  \city{Beijing}
  \country{China}
  }
\email{dmjin@stu.pku.edu.cn}

\author{Lei Li}
\affiliation{%
   \institution{The University of Hong Kong}
  \city{Hong Kong}
  \country{China}
  }
\email{nlp.lilei@gmail.com}

\author{Zhi Jin}
\authornote{Corresponding author.}
\affiliation{%
   \institution{Wuhan University, \\ Peking University}
  \city{Wuhan}
  \country{China}
  }
\email{zhijin@whu.edu.cn}

\renewcommand{\shortauthors}{Trovato et al.}

\begin{abstract}
Code generation refers to automatically producing executable programs from user requirements. Recently, researchers have explored approaches to enhance the correctness of generated code with advanced large language models. Although achieving improvements, existing approaches focus on designing reasoning strategies or post-refinement methods to enhance code generation performance. Despite their differences, all these methods share a common assumption: the LLM can correctly understand the given requirement. However, this assumption does not always hold. To fill this gap, we propose REA-Coder, a requirement alignment approach to enhance the code generation performance of LLMs. REA-Coder involves first identifying the requirement content that does not align with LLMs and aligning the requirements. Then, based on the aligned requirements, LLMs generate code and further verify whether the generated code aligns with the requirements, iterating this process of requirement alignment and code generation until generating correct code or achieving the maximum number of iterations. Experimental results show that REA-Coder outperforms all advanced baselines on four LLMs across five programming benchmarks. Concretely, REA-Coder achieves average improvements of 7.93\%, 30.25\%, 26.75\%, 8.59\%, and 8.64\% on the five benchmark datasets, demonstrating the effectiveness of requirement alignment for improving the code generation performance of LLMs. 
\end{abstract}



\received{20 February 2007}
\received[revised]{12 March 2009}
\received[accepted]{5 June 2009}

\maketitle

\input{1-Introduction}

\input{3-Approach}
\input{4-Experiments}

\input{5-Experimental_Results}
\input{6-Analysis}

\input{2-Related_Work}

\input{7-Conclusion}


\bibliographystyle{ACM-Reference-Format}
\bibliography{main}


\end{document}

%% file: 1-Introduction.tex
\section{Introduction}

Code generation aims to automatically transform user requirements into executable programs. With the rapid advances in large language models (LLMs), automated code generation has attracted significant attention from both academia and industry~\cite{li2023codeeditor, islam2025codesim, zhang2024codeagent}. Despite notable progress, LLMs still struggle with complex requirements. For example, Qwen3-Coder~\cite{cao2026qwen3} achieves only 15.15\% Pass@1 on competitive programming benchmark CodeContests~\cite{li2022competition}. This limits the application of LLMs in practical software development.

To date, various methods have been proposed to improve code generation. One category of work designs reasoning strategies to enhance the generation process, such as structured chain-of-thought prompting~\cite{li2025structured}, difficulty-aware routing~\cite{li2025intention}, and iterative plan-code-test workflows~\cite{zhang2024pair}. Another category focuses on post-processing to fix the generated code. Some methods directly repair code based on execution feedback~\cite{olausson2023self,tian2025fixing}, while others~\cite{mu2023clarifygpt,tian2025aligning,jia2025automated} go further by detecting the inconsistency between generated code and the requirement, and refining code accordingly.

Despite their differences, all these methods share a common assumption: the LLM can correctly understand the given requirement. Reasoning-based methods focus on improving how the model plans and generates code. Post-processing methods focus on repairing outputs or refining requirements based on the model's generated results. However, none of them verifies whether the model has truly understood what the requirement asks for.

\begin{figure}[htbp]
  \centering 
  \includegraphics[width=\linewidth]{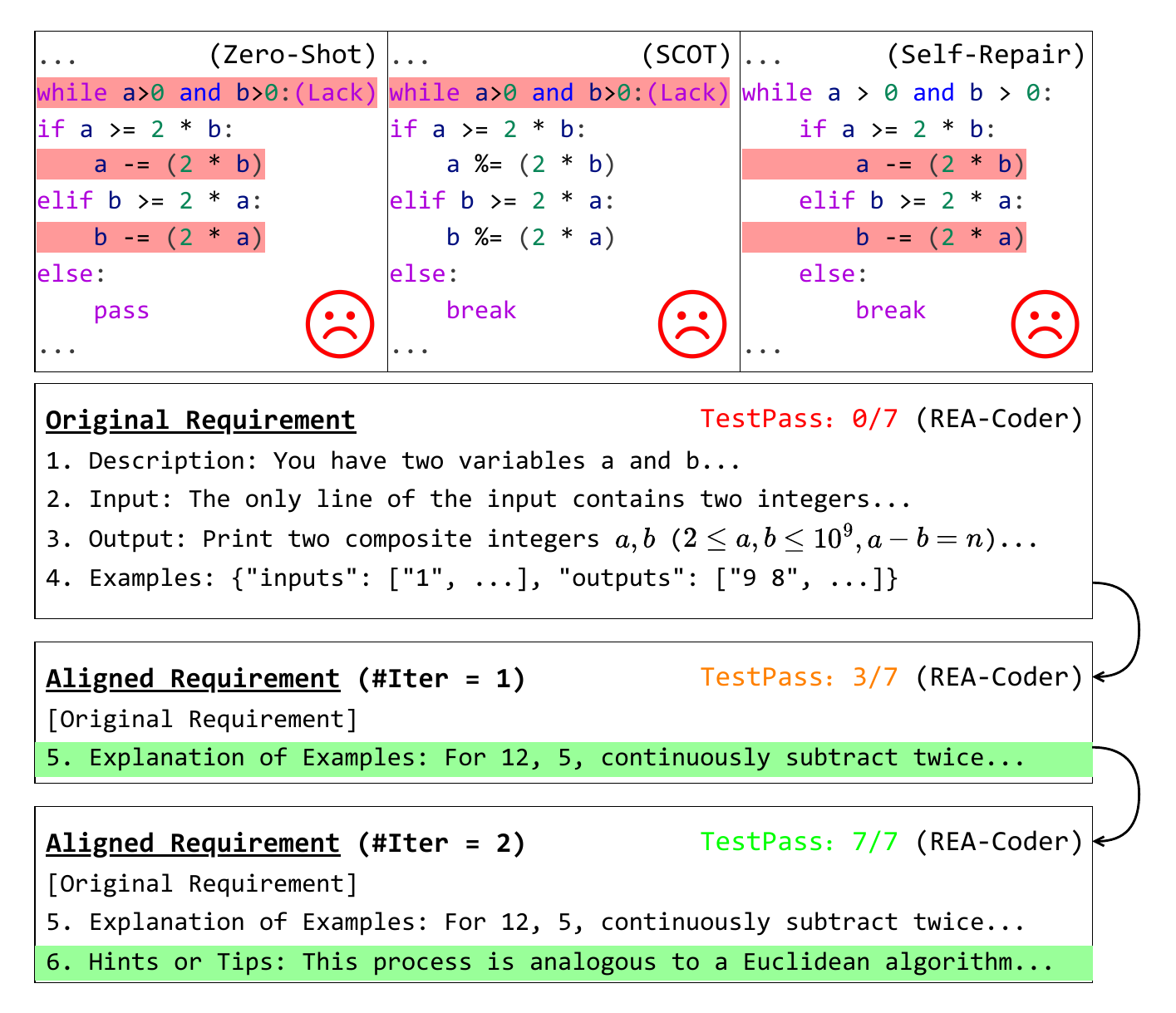}  
  \vspace{-6mm}
  \caption{An example of how requirement alignment improves code generation over iterations.}  
  \label{fig:iter_case}  
  \vspace{-4mm}
\end{figure}

Our analysis reveals that this assumption does not always hold (as illustrated in Figure~\ref{fig:iter_case}). LLMs can fundamentally misunderstand the requirement, rather than merely making mistakes during reasoning or code writing. When such misunderstanding occurs, the model is working from a wrong interpretation of the task. In this case, neither stronger reasoning strategies nor output-level repairs can fix the root cause, as they are all built upon the same flawed understanding.

To address this limitation, we propose REA-Coder, a requirement alignment approach designed to boost the code generation performance of LLMs. REA-Coder starts by identifying the requirement content that does not align with LLMs, where we construct requirement-oriented question checklists and reference answers that cover core requirement elements. By comparing the model’s generated answers of questions against the reference answers, REA-Coder pinpoints requirement items misunderstood by LLMs, and further aligns the requirement.
Then, based on the aligned requirement, LLMs generate code and further verify whether the generated code aligns with the requirement, iterating this process of requirement alignment and code generation until the generated code passes all public test cases or hits a preset iteration budget.
In this way, REA-Coder enables LLMs to generate functionally correct code based on requirement alignment.

We conduct experiments to evaluate the effectiveness of REA-Coder on four LLMs (\ie DeepSeek-v3.2~\cite{liu2025deepseek}, Qwen3-Coder~\cite{cao2026qwen3}, GPT-5-mini~\cite{singh2025openai}, and Gemini-3-Flash~\cite{gemini3flash_announcement}) with five widely used programming benchmarks(\ie APPS~\cite{hendrycks2021measuring}, CodeContests-raw~\cite{li2022competition}, CodeContests~\cite{li2022competition}, xCodeEval~\cite{khan2024xcodeeval}, and LiveCodeBench-Lite~\cite{jain2025livecodebench}). Experimental results show that REA-Coder significantly outperforms 8 state-of-the-art baselines across 20 (4 $\times$ 5) model and benchmark combinations, validating the significant effectiveness of aligning requirements with LLMs in code generation. 
Concretely, REA-Coder achieves average improvements of 7.93\%, 30.25\%, 26.75\%, 8.59\%, and 8.64\% on the five benchmarks, respectively.
Ablation studies show that both aligning the requirement before code generation and alignment requirement verification after generating code make substantial contributions to the effectiveness of REA-Coder.
We also investigate the effect of the number of iterations. We find that REA-Coder outperforms other iterative approaches in all iteration times.
In addition, the performance gain is much larger in the early iterations, and then gradually becomes smaller, which suggests that requirement misalignment is particularly prominent during the early iterations. 
We also verify the first generated code based only on requirement alignment before code generation outperforms the zero-shot approach 210.44\% and 344.67\% on APPS and xCodeEval. 
The results validate the central design principle of REA-Coder that requirement alignment should be moved as early as possible, rather than relying primarily on execution feedback after acquiring code.

The main contributions of this paper are summarized as follows:
\begin{itemize}
    \item  We argue that the assumption (\ie the LLM can correctly understand the given requirement) does not always hold.
    \item  We propose REA-Coder, an approach to enhance the code generation performance of LLMs through sophisticated requirement alignment. 
    \item  We evaluate REA-Coder across four LLMs and five programming benchmarks, achieving state-of-the-art performance compared to 8 advanced code generation baselines on all (4 $\times$ 5) LLMs and benchmarks.
\end{itemize}

\section{Motivation Example}

Figure \ref{fig:iter_case} presents an example from APPS. In this example, input two integers $a$ and $b$, the requirement is to repeat the following process: if $a = 0$ or $b = 0$, or both $a < 2b$ and $b < 2a$, the process ends. Otherwise, if $a \ge 2b$, subtract $2b$ from $a$ repeatedly until $a < 2b$; else if $b \ge 2a$, subtract $2a$ from $b$ repeatedly until $b < 2a$. Finally, output the resulting values of $a$ and $b$.

We first employ Qwen3-Coder with zero-shot prompting to directly generate code based on the requirement. However, the generated code is incorrect. Specifically, the LLM does not implement the required loop and uses subtraction to execute the process, which leads to time limit exceeded. This misunderstanding leads to an incorrect implementation that fails to satisfy the output constraint.

To further investigate whether existing approaches can resolve this issue, we apply reasoning-based and post-processing methods (\eg SCOT~\cite{li2025structured} and Self-Repair~\cite{olausson2023self}). However, they still fail to correct the core logic (one fails to implement the required loop, while the other does not use the modulo operation to replace repeated subtraction). This is because all subsequent reasoning and repair steps are built upon the same flawed understanding of the requirement.

In contrast, REA-Coder explicitly aligns the LLM with the requirement before code generation. Through the requirement alignment stage, REA-Coder is able to identify a critical misunderstanding and iteratively align the requirement. In subsequent iterations, it explicitly provides explanations of input-output examples and further introduces additional guidance, such as ``use the Euclidean algorithm for solving''.
With the aligned requirement, the model is finally able to generate correct code that passes all test cases. 
This example highlights the key challenges. LLMs may fundamentally misunderstand the requirement, rather than merely making implementation errors. Reasoning and repair methods are insufficient to address errors caused by requirement misunderstanding. 
These observations motivate the design of REA-Coder, which systematically performs requirement alignment before code generation and alignment verification after generation.

%% file: 3-Approach.tex
\begin{figure*}[htbp]
  \centering 
  \includegraphics[width=\linewidth]{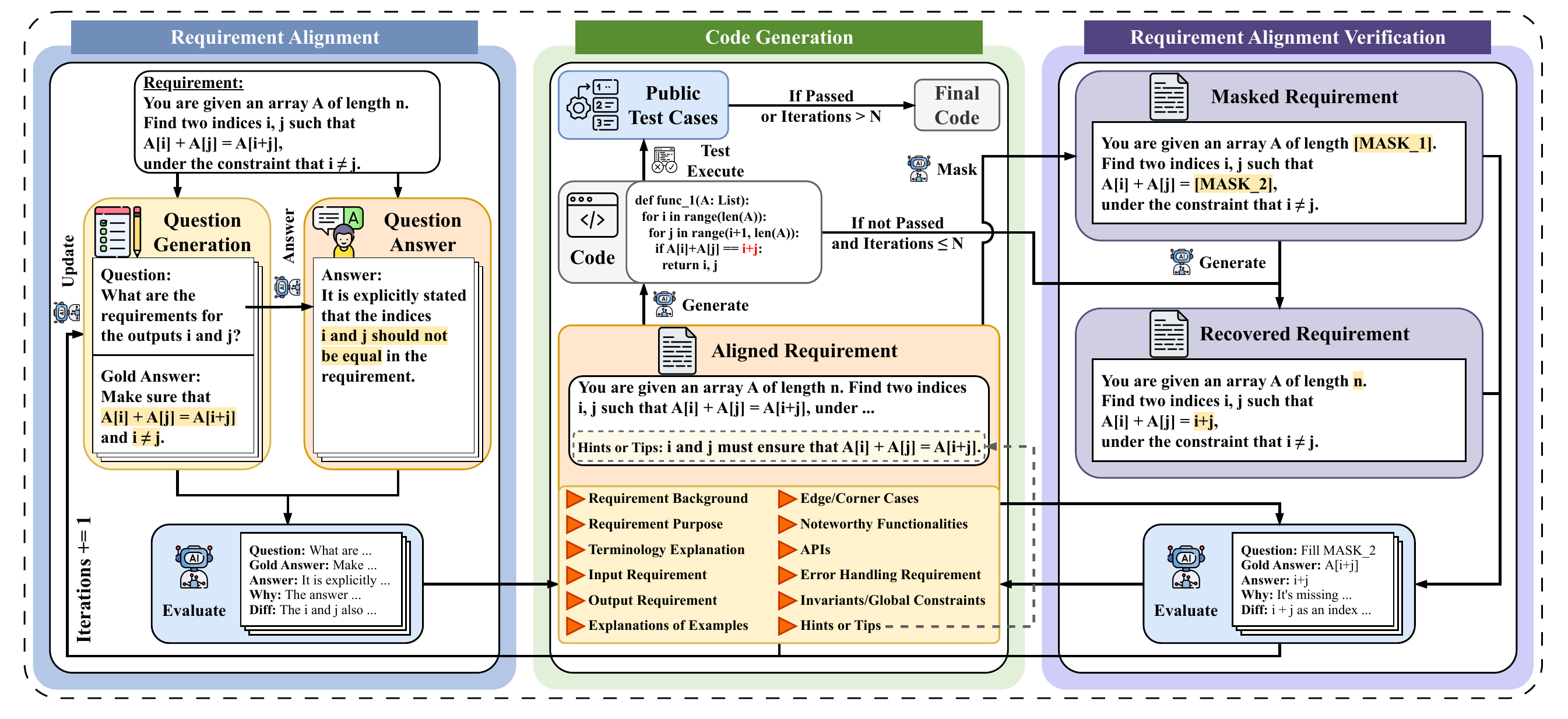}  
  \vspace{-6mm}
  \caption{Overview of REA-Coder.} 
  \label{fig:overview}  
  \vspace{-3mm}
\end{figure*}

\begin{figure}[htbp]
  \centering 
  \includegraphics[width=\linewidth]{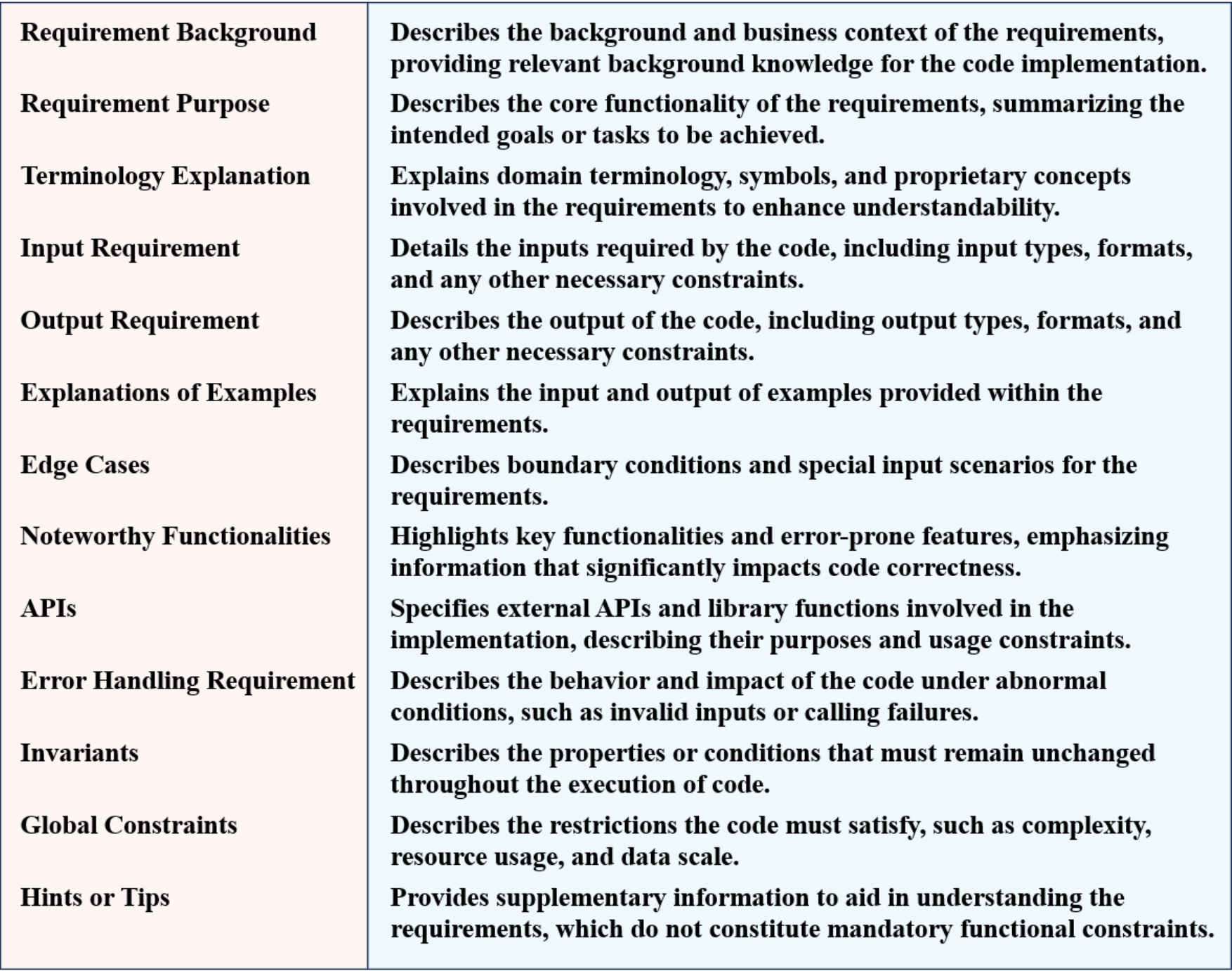} 
  \vspace{-4mm}
  \caption{Core requirement dimensions in REA-Coder.} 
  \label{fig:explanation}  
  \vspace{-4mm}
\end{figure}

\section{Approach}
In this paper, we propose REA-Coder, a requirement alignment approach to improve the code generation performance of LLMs. 
As illustrated in Figure \ref{fig:overview}, REA-Coder first examines the model's comprehension of the requirement from multiple dimensions, and then aligns requirements with the model.
Based on the aligned requirement, the model begins to generate code. If the generated code fails to pass the public test cases, we mask key semantic spans of the requirement and task the LLM to recover the masked spans based on the generated code. The intuition is that if the code is well-aligned with the original requirement, the masked spans can be correctly recovered; otherwise, recovery errors indicate misalignment. The recovered content is therefore compared against the original unmasked requirement to identify and correct misalignment, after which the entire process of requirement alignment and code generation is restarted. This process is repeated iteratively until the code passes all public test cases or the predefined maximum number of iterations is reached.

\subsection{Requirement Alignment}  \label{sec: sec 3.1}
Direct code generation immediately upon requirement input may lead to fundamental misinterpretation of the requirement by the LLM as demonstrated in Figure \ref{fig:iter_case}, resulting in erroneous outputs across all subsequent stages. To prevent the model from generating code based on a misaligned requirement, REA-Coder verifies the model’s understanding of the input requirement. To render the process of requirement understanding explicit and testable, we define a set of requirement dimensions. These dimensions build on established foundations in requirements engineering \cite{glinz2000uml,greenspan1994rml} and align with the IEEE recommended practice for software requirements specifications \cite{ieee2011srs}. The core requirement dimensions are shown in Figure \ref{fig:explanation}. Based on these dimensions, REA-Coder first requires the LLM to analyze the original requirement and construct a question checklist along with corresponding reference answers from the core requirement dimensions. To guide the LLM to generate high-quality reference answers, we add a small set of manually annotated few-shot examples as additional input in the prompt. These examples are high-quality question-answer pairs for typical code generation tasks that are strictly aligned with the requirements, which clarify the expected format, specificity requirements of the generated content, and the criteria for strict alignment with the original requirement text.

After acquiring the question checklist, REA-Coder is instructed to answer the constructed questions based on the current requirement. The LLM then compares REA-Coder’s answers with the reference answers, and identifies whether the LLM correctly understands the requirement for each question. For correctly answered questions, which indicate that the model has properly comprehended the relevant content, they are removed from the question checklist. For every incorrectly answered question, REA-Coder records the question itself, the reference answer, and the model’s generated answer, and produces corresponding feedback. 

Upon receiving the evaluation feedback on requirement understanding deviations, REA-Coder conducts systematic integration and attribution analysis of the identified misunderstandings. Specifically, it uses the LLM to map the evaluation feedback back to the predefined system of core requirement dimensions, and on this basis, constructs an aligned requirement with disambiguated and supplemented information, as illustrated in Figure~\ref{fig:explanation}. The aligned requirement fully retains the original problem description, while explicitly supplementing the requirement elements that the LLMs failed to correctly align.

\subsection{Code Generation}
Upon completion of the requirement alignment, REA-Coder performs code generation based on the enhanced requirement. Most existing code generation paradigms directly take raw, unprocessed original requirements as input, while REA-Coder performs code generation with the aligned requirement as its core input. Accordingly, REA-Coder is not limited to improving generation performance via existing approaches such as designing reasoning strategies or post-processing of the generated code. 
Instead, it adheres to the core principle of aligning LLMs with requirements first, and code generation second: after correcting the model’s misunderstanding of the requirements, it performs code generation based on the requirement that is fully aligned with the target intent.

For the generated code, we perform functional correctness verification via public test cases. If the generated code passes all public test cases, the iterative optimization process terminates. The current code is returned as the final output. 
If the code fails to pass all public test cases, REA-Coder enters the subsequent in-depth requirement alignment process as described in Section \ref{section 2.3}.

\subsection{Requirement Alignment Verification} \label{section 2.3}
Although the question answering-based requirement alignment and enhancement mechanism can bridge the gap between the model’s comprehension and the user’s intent, they are still insufficient to guarantee the deep logical consistency between the generated code and the target requirement. The model may fail to internalize the requirement semantics into the program logic it generates, leading to potential functional deviations in the code. To further verify the authenticity and depth of requirement alignment after code generation, REA-Coder introduces a masked-requirement-recovery verification mechanism.

In the requirement alignment verification stage, REA-Coder first masks a portion of the key semantic spans of the requirement. To ensure that the masking strategy can effectively probe the misunderstanding points of generated code while maintaining the contextual semantic coherence and derivability, we design the following two core masking rules.

The first rule adheres to a content priority principle. 
We focus masking operations on the semantic elements within the core requirement dimensions defined in Section \ref{sec: sec 3.1} since they play a decisive role in code logic generation.
The second rule follows a semantic coherence principle, which is designed to ensure the masked requirement remains semantically interpretable and logically complete. To this end, we set the following executable rules for masking operations: (1) Each masked span covers a complete semantic unit (\eg noun phrases, verb phrases, constraint clauses) without splitting individual words; (2) Sufficient consecutive unmasked context must be reserved between any two adjacent masked spans to guarantee contextual semantic continuity; (3) The number of masked spans in a sentence is limited to preserve the logical consistency of each sentence; (4) Masking operations shall not damage the overall framework of the requirement, where the suggestive keywords (\eg input, output, and constraint) must be retained meanwhile their values of these keys can be masked.

Next, REA-Coder requires LLM to recover the masked content based on the generated code. 
If the generated code faithfully embodies the logic of the requirement, the LLM should be able to accurately infer the masked requirement spans according to the generated code. Conversely, systematic deviations between the recovered content and the requirement indicate that the generated code is still based on an incomplete requirement understanding of LLMs.

To verify the consistency, REA-Coder conducts a semantic comparison across multiple dimensions between the original enhanced requirement and the requirement content recovered by the model. For each masked span, the model generates corresponding feedback covering the accuracy judgment of the recovery, the discrepancy analysis between the recovered content and the original semantics, and the core requirement semantic elements that remain missing. This feedback is converted into complementary requirement alignment signals and fed back to the requirement alignment stage.

Finally, REA-Coder updates the question checklist for the requirement alignment stage. It removes questions that have passed verification, retains previously incorrectly answered questions, and converts the newly exposed requirement understanding deviations from the requirement alignment verification stage into new question items. The iterative process is restarted based on the updated question checklist. The iteration terminates only when the generated code passes all public test cases, or the predefined maximum number of iterations is reached.

%% file: 4-Experiments.tex
\section{Experiments}

We aim to answer the following research questions (RQs):
\begin{itemize}
    \item  RQ1: What is the performance of REA-Coder in code generation tasks compared to the state-of-the-art approaches?
    \item  RQ2: What is the contribution of each component in REA-Coder to the overall effectiveness?
    \item  RQ3: How does the number of iterations influence the performance of REA-Coder?
\end{itemize}

\subsection{Benchmarks}

To comprehensively evaluate the performance of our approach, we apply five code generation benchmarks, including APPS~\cite{hendrycks2021measuring}, CodeContests-raw~\cite{li2022competition}, CodeContests~\cite{li2022competition}, xCodeEval~\cite{khan2024xcodeeval}, and LiveCodeBench-Lite~\cite{jain2025livecodebench}. These benchmarks are widely used, providing a robust testbed for assessing code generation capabilities of models.

\textbf{APPS} \cite{hendrycks2021measuring} aggregates programming problems from various online coding platforms (\eg Codeforces, LeetCode) across diverse difficulty levels. It contains 5,000 training data and 5,000 test data. Each example provides requirement descriptions, test cases, and reference solutions. Following existing works \cite{tian2025aligning}, we use 300 problems of the test set to balance evaluation costs and statistical representativeness, which proportionally matches its original difficulty distribution (\ie introductory, interview, and competition). 

\textbf{CodeContests-raw} \cite{li2022competition} is introduced by Google DeepMind, which is sourced from platforms like Codeforces and AtCoder. This benchmark is designed to evaluate models on competitive programming problems. CodeContests \cite{li2022competition} extends CodeContests-raw by increasing the number of hidden test cases per problem to enforce stricter evaluation criteria. Following the evaluation established by previous works \cite{chen2022codet, zhang2023planning, tian2025aligning}, we use 165 problems from both CodeContests-raw and CodeContests, respectively.

\textbf{xCodeEval} \cite{khan2024xcodeeval} is a large-scale, competition-level code generation benchmark containing approximately 7,500 programming problems. It divides these problems into three levels, including easy, medium, and hard modes. 
Similarly, we use a subset of data with 300 problems to evaluate our approach and baselines following the previous works \cite{tian2025aligning}.

\textbf{LiveCodeBench-Lite} \cite{jain2025livecodebench} comprises recent programming problems sourced from real-world online judge environments. Compared with static benchmarks built from older datasets, it better reflects the evolving difficulty, style, and implementation requirements of contemporary code generation. We include this benchmark to further validate the practical generalization of our approach in real-world code generation scenarios.

\subsection{Comparison Baselines}
We benchmark our approach against a diverse set of representative code generation baselines. Beyond the standard zero-shot baseline, the compared methods can be broadly categorized into two paradigms. Reasoning-based paradigm: SCoT \cite{li2025structured}, Self-Collaboration \cite{dong2024self}, ICOT \cite{li2025intention}, which introduces reasoning strategies to enhance the quality of generated code. Post-processing paradigm: Self-Repair~\cite{olausson2023self}, SpecFix \cite{jia2025automated}, $\mu$Fix \cite{tian2025fixing}, and Specine \cite{tian2025aligning}, aiming to fix generated code through execution feedback or detecting semantic misalignment between requirements and output code.

\subsection{Base LLMs}
To evaluate the effectiveness of REA-Coder, we conduct experiments on four advanced LLMs (\ie DeepSeek-v3.2-Chat \cite{liu2025deepseek}, Qwen3-Coder-30B-A3B-Instruct \cite{cao2026qwen3}, GPT-5-mini \cite{singh2025openai}, and Gemini-3-Flash-Preview \cite{gemini3flash_announcement}) across five benchmarks compared to eight baselines.

\subsection{Metrics}
Following previous works, we use Pass@k, a popular metric in code generation, as shown in Format \ref{format: metrics}. It is the expectation of passing all tests of a task at least once within $k$ attempts, where $n$ is the number of solutions of a task sampled from an LLM and $c$ is the number of correct solutions. In this paper, we use Pass @1 since in real-world scenarios, developers usually only consider the single generated code.
\begin{equation}
\label{format: metrics}
\text{Pass@k} = {{E}}\left[1 - \frac{\binom{n - c}{k}}{\binom{n}{k}}\right]
\end{equation}

We further evaluate the efficiency of REA-Coder in terms of time overhead and token overhead. Time overhead measures the average runtime of the method, and token overhead measures the average tokens consumed, including both prompt tokens and generated tokens. Lower values of time overhead and token overhead indicate better efficiency.

\subsection{Implementation Details}
To improve generation stability, we set the temperature to 0 for code generation and 0.2 for other steps. The maximum number of questions generated is 20, and the maximum number of iterations is set to 10. Moreover, public test cases are used during the iterative process to assess whether the generated code satisfies the requirement, whereas the final results are reported only on hidden test cases in order to avoid evaluation bias and potential leakage.

%% file: 5-Experimental_Results.tex
\section{Experimental Results}

\begin{table*}[t]
\centering
\caption{Overall performance and efficiency across LLMs and benchmarks.}
\vspace{-3mm}
  \resizebox{0.95\linewidth}{!}{
\small
\begin{tabular}{llccccccc}
\toprule

\multirow{2}{*}{LLM} &
\multirow{2}{*}{Method} &
\multicolumn{5}{c}{Pass@1} &
\multirow{2}{*}{Time (h)} &
\multirow{2}{*}{Token (M)} \\

\cmidrule(lr){3-7}

& & APPS & CodeContests-raw & CodeContests & xCodeEval & LC-Lite & & \\

\midrule

\multirow{10}{*}{DeepSeek-v3.2}
& zero-shot & 66.00\% & 38.79\% & 37.20\% & 50.33\% & 40.57\% & 0.32 & 0.37\\
& SCoT~\cite{li2025structured} & 67.33\% & 41.21\% & 38.18\% & 53.67\% & 52.00\% & 0.47 & 1.20\\
& ICoT~\cite{li2025intention} & 68.00\% & 47.27\% & 43.03\% & 59.00\% & 52.57\% & 0.62 & 0.96\\
& Self-Collaboration~\cite{dong2024self} & 70.33\% & 42.42\% & 39.39\% & 60.33\% & 50.86\% & 2.56 & 6.60\\
& $\mu$Fix~\cite{tian2025fixing} & 71.67\% & 53.94\% & 49.70\% & 63.33\% & 56.57\% & 1.14 & 1.66\\
& Self-Repair~\cite{olausson2023self} & 71.00\% & 52.72\% & 46.67\% & 62.33\% & 50.86\% & 0.68 & 1.45\\
& SpecFix~\cite{jia2025automated} & 72.00\% & 47.27\% & 44.24\% & 63.33\% & 54.86\% & 3.08 & 13.61\\
& Specine~\cite{tian2025aligning} & 74.33\% & 50.30\% & 48.48\% & 65.00\% & 56.57\% & 2.32 & 5.48\\
& \textbf{REA-Coder} & \textbf{81.67\%} & \textbf{67.27\%} & \textbf{61.82\%} & \textbf{70.33\%} & \textbf{62.86\%} & 2.78 & 9.74\\
&  & \cellcolor{tablegreen!80} \textbf{9.87\%} $\uparrow$  & \cellcolor{tablegreen!80} \textbf{24.71\%} $\uparrow$ & \cellcolor{tablegreen!80} \textbf{24.39\%} $\uparrow$ & \cellcolor{tablegreen!80} \textbf{8.20\%} $\uparrow$ & \cellcolor{tablegreen!80} \textbf{11.12\%} $\uparrow$ & -- & -- \\
\midrule

\multirow{10}{*}{Qwen3-Coder}
& zero-shot & 16.00\% & 16.97\% & 15.15\% & 9.67\% & 16.57\% & 0.34 & 0.74\\
& SCoT & 37.33\% & 19.39\% & 13.33\% & 26.33\% & 30.29\% & 0.51 & 1.79\\
& ICoT & 44.00\% & 15.76\% & 12.73\% & 28.33\% & 31.43\% & 0.32 & 0.90\\
& Self-Collaboration & 40.33\% & 18.79\% & 13.93\% & 33.00\% & 28.00\% & 2.44 & 8.77\\
& $\mu$Fix & 43.67\% & 20.61\% & 15.15\% & 29.33\% & 33.14\% & 1.22 & 2.94\\
& Self-Repair & 49.33\% & 22.56\% & 15.85\% & 27.67\% & 33.71\% & 0.72 & 1.89\\
& SpecFix & 56.67\% & 22.42\% & 19.39\% & 44.67\% & 33.14\% & 2.78 & 13.92\\
& Specine & 52.67\% & 23.03\% & 20.00\% & 44.00\% & 32.00\% & 1.97 & 5.56\\
& \textbf{REA-Coder} & \textbf{66.67}\% & \textbf{40.61}\% & \textbf{33.33}\% & \textbf{52.00}\% & \textbf{38.29}\% & 2.65 & 11.78\\
& & \cellcolor{tablegreen!80} \textbf{17.68}\% $\uparrow$ & \cellcolor{tablegreen!80} \textbf{76.34}\% $\uparrow$ & \cellcolor{tablegreen!80} \textbf{66.65}\% $\uparrow$ & \cellcolor{tablegreen!80} \textbf{16.41}\% $\uparrow$ & \cellcolor{tablegreen!80} \textbf{15.54}\% $\uparrow$ & -- & -- \\
\midrule

\multirow{10}{*}{GPT-5-mini}
& zero-shot & 79.33\% & 61.82\% & 45.45\% & 54.67\% & 48.00\% & 0.18 & 0.34\\
& SCoT & 73.67\% & 48.48\% & 44.24\% & 56.00\% & 56.57\% & 0.22 & 0.79\\
& ICoT & 67.67\% & 43.03\% & 40.61\% & 51.33\% & 53.14\% & 0.67 & 0.80\\
& Self-Collaboration & 78.00\% & 66.06\% & 57.58\% & 57.67\% & 54.86\% & 1.85 & 3.40\\
& $\mu$Fix & 75.33\% & 51.52\% & 49.09\% & 58.67\% & 52.57\% & 0.77 & 1.15\\
& Self-Repair & 73.67\% & 49.09\% & 45.45\% & 55.33\% & 56.57\% & 0.54 & 0.89\\
& SpecFix & 79.33\% & 52.73\% & 50.91\% & 60.33\% & 57.71\% & 2.92 & 9.79\\
& Specine & 82.89\% & 63.03\% & 55.76\% & 69.00\% & 60.00\% & 1.82 & 7.46\\
& \textbf{REA-Coder} & \textbf{85.00}\% & \textbf{73.33}\% & \textbf{60.61}\% & \textbf{71.33}\% & \textbf{62.86}\% & 2.36 & 8.58\\
& & \cellcolor{tablegreen!80} \textbf{2.55}\% $\uparrow$ & \cellcolor{tablegreen!80} \textbf{11.01}\% $\uparrow$ & \cellcolor{tablegreen!80} \textbf{5.26}\% $\uparrow$ & \cellcolor{tablegreen!80} \textbf{3.38}\% $\uparrow$ & \cellcolor{tablegreen!80} \textbf{4.77}\% $\uparrow$ & -- & -- \\
\midrule

\multirow{10}{*}{Gemini-3-Flash}
& zero-shot & 79.00\% & 55.76\% & 53.33\% & 69.67\% & 57.14\% & 0.12 & 0.38\\
& SCoT & 80.00\% & 58.79\% & 56.36\% & 73.33\% & 69.14\% & 0.15 & 0.71\\
& ICoT & 81.33\% & 61.21\% & 58.18\% & 73.67\% & 70.29\% & 0.44 & 0.79\\
& Self-Collaboration & 81.67\% & 66.67\% & 63.03\% & 75.33\% & 68.00\% & 1.27 & 3.21\\
& $\mu$Fix & 82.67\% & 69.09\% & 63.03\% & 76.33\% & 72.00\% & 0.65 & 1.82\\
& Self-Repair & 82.00\% & 68.48\% & 64.85\% & 74.00\% & 71.43\% & 0.81 & 1.66\\
& SpecFix & 80.00\% & 67.88\% & 63.63\% & 75.67\% & 73.14\% & 2.51 & 9.95\\
& Specine & 86.91\% & 74.55\% & 67.88\% & 78.33\% & 71.43\% & 1.33 & 7.33\\
& \textbf{REA-Coder} & \textbf{88.33}\% & \textbf{81.21}\% & \textbf{75.15}\% & \textbf{83.33}\% & \textbf{75.43}\% & 1.98 & 8.96\\
& & \cellcolor{tablegreen!80} \textbf{1.63}\% $\uparrow$ & \cellcolor{tablegreen!80} \textbf{8.93}\%$\uparrow$  & \cellcolor{tablegreen!80} \textbf{10.71}\% $\uparrow$ & \cellcolor{tablegreen!80} \textbf{6.38}\% $\uparrow$ & \cellcolor{tablegreen!80} \textbf{3.13}\% $\uparrow$ & -- & -- \\
\bottomrule
\end{tabular}
}

\label{tab:pass1_results}

\end{table*}

\begin{table*}[t]
\caption{Ablation results across LLMs and benchmarks.}
\centering
\vspace{-3mm}
\resizebox{0.83\linewidth}{!}{
\small
\begin{tabular}{llccccc}
\toprule
\multirow{2}{*}{LLM} & \multirow{2}{*}{Method} & \multicolumn{5}{c}{Pass@1} \\
\cmidrule(lr){3-7}
& & APPS & CodeContests-raw & CodeContests & xCodeEval & LiveCodeBench-Lite \\
\midrule
\multirow{3}{*}{DeepSeek-v3.2}
& \textbf{REA-Coder} & \textbf{81.67\%} & \textbf{67.27\%} & \textbf{61.82\%} & \textbf{70.33\%} & \textbf{62.86\%} \\
& $ REA\text{-}Coder_{WO-QA}$ & 81.33\% & 63.03\% & 58.79\% & 67.67\% & 57.71\% \\
& $ REA\text{-}Coder_{WO-MASK}$ & 80.33\% & 58.18\% & 54.54\% & 69.67\% & 57.14\% \\
\midrule
\multirow{3}{*}{Qwen3-Coder}
& \textbf{REA-Coder} & \textbf{66.67\%} & \textbf{40.61\%} & \textbf{33.33\%} & \textbf{52.00\%} & \textbf{38.29\%} \\
& $ REA\text{-}Coder_{WO-QA}$ & 63.00\% & 36.97\% & 30.30\% & 50.00\% & 35.43\% \\
& $ REA\text{-}Coder_{WO-MASK}$ & 59.00\% & 29.70\% & 25.45\% & 45.67\% & 33.14\% \\
\midrule
\multirow{3}{*}{GPT-5-mini}
& \textbf{REA-Coder} & \textbf{85.00\%} & \textbf{73.33\%} & \textbf{60.61\%} & \textbf{71.33\%} & \textbf{62.86\%} \\
& $ REA\text{-}Coder_{WO-QA}$ & 83.33\% & 65.14\% & 54.86\% & 70.00\% & 58.86\% \\
& $ REA\text{-}Coder_{WO-MASK}$ & 83.33\% & 67.27\% & 56.36\% & 66.33\% & 61.14\% \\
\midrule
\multirow{3}{*}{Gemini-3-Flash}
& \textbf{REA-Coder} & \textbf{88.33\%} & \textbf{81.21\%} & \textbf{75.15\%} & \textbf{83.33\%} & \textbf{75.43\%} \\
& $ REA\text{-}Coder_{WO-QA}$ & 86.00\% & 78.79\% & 72.12\% & 80.00\% & 71.43\% \\
& $ REA\text{-}Coder_{WO-MASK}$ & 86.67\% & 75.76\% & 71.52\% & 80.00\% & 74.86\% \\
\bottomrule
\end{tabular}
}

\label{tab:pass1_no_cost}
\end{table*}

\begin{figure*}[htbp]
  \centering 
  \includegraphics[width=\linewidth]{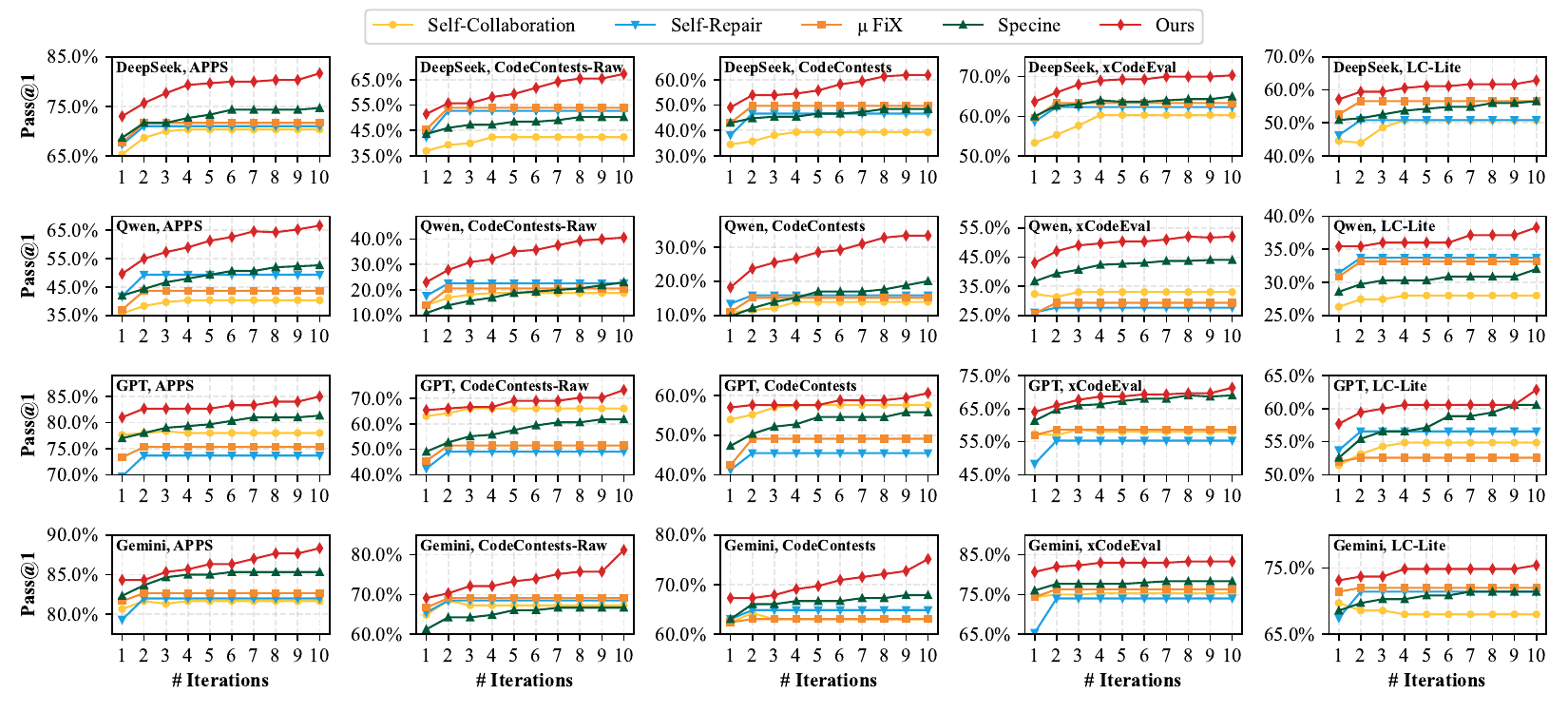}  
  \vspace{-7mm}
  \caption{Pass@1 across iterations for REA-Coder and iterative baselines.} 
  \label{fig:iteration_ablation}  
  \vspace{-2mm}
\end{figure*}

\subsection{RQ1: Overall Performance}

To answer RQ1, we compare REA-Coder with zero-shot and a diverse set of representative baselines, including reasoning-based methods and post-processing methods. We conduct experiments on four LLMs across five benchmarks, using Pass@1 as the evaluation metric, and additionally report time cost and token cost to assess the effectiveness-efficiency trade-off of REA-Coder in a more comprehensive manner.
Table~\ref{tab:pass1_results} shows the performance of all methods across different models and benchmarks.

Table~\ref{tab:pass1_results} shows that in experiments across all four LLMs and five benchmarks, REA-Coder consistently maintains the optimal Pass@1 performance when compared with eight baselines. Compared with the existing state-of-the-art methods, REA-Coder achieves average improvements of 7.93\%, 30.25\%, 26.75\%, 8.59\%, and 8.64\% on the five benchmarks, respectively. Grouped by methodological paradigm, REA-Coder outperforms the four types of reasoning-based methods by an average of 30.34\%, and the four types of post-processing methods by an average of 19.51\%. When requirements are not properly aligned, even advanced reasoning methods and post-processing strategies can still lead to incorrect code generation. In contrast, REA-Coder verifies requirement understanding before code generation, which helps reduce errors caused by misunderstood requirements.

REA-Coder achieves a more pronounced improvement on more challenging benchmarks. Among the five benchmarks in our work, APPS is a relatively easy dataset with routine programming tasks, while the remaining four benchmarks consist of competitive programming problems with higher task complexity. On CodeContests-raw and CodeContests, REA-Coder achieves improvements of 30.25\% and 26.75\% over the existing state-of-the-art methods, which are significantly higher than its performance on the APPS (7.93\% relative improvements). This result indicates that REA-Coder delivers a particularly prominent optimization effect when the requirements involve more complex constraints, where requirement misunderstanding is more likely to occur.

In addition, the gains are also larger on relatively weaker models. The average relative improvement of REA-Coder reaches 38.13\% on Qwen3-Coder and 15.66\% on DeepSeek-v3.2, while the corresponding improvements on Gemini-3-Flash and GPT-5-mini are 6.16\% and 5.39\%. This result shows that REA-Coder brings larger improvements on models with weaker code generation ability. A likely reason is that these models cannot fully align with the requirement during code generation, and therefore benefit more from the requirement alignment process in REA-Coder. At the same time, the results also show that requirement alignment remains beneficial even when the model already has strong code generation ability.

In terms of efficiency, REA-Coder requires more time and tokens than zero-shot and reasoning-based methods, which is expected given its multi-stage requirement alignment process. However, compared with other post-processing methods through requirement alignment such as SpecFix and Specine, REA-Coder has a comparable cost. Meanwhile, REA-Coder achieves consistently higher Pass@1 scores across all settings, which makes the additional cost over simpler baselines a worthwhile trade-off.

\vspace{1mm}
\begin{custommdframed}
\textbf{Finding 1:} REA-Coder consistently achieves the best overall code generation performance across different LLMs and benchmarks. Its gains are more pronounced on relatively weaker models and more challenging benchmarks.
\end{custommdframed}
\vspace{0mm}

\subsection{RQ2: Contribution of Main Component}
To answer RQ2, we perform an ablation study by removing each of the two main stages in REA-Coder. Specifically, $REA\text{-}Coder_{WO-QA}$ removes the QA-based requirement alignment stage and retains only the code generation and masking-based verification stages. $REA\text{-}Coder_{WO-MASK}$ removes the masking-based verification stage and retains only the requirement alignment and code generation stages. Note that both variants retain the iterative refinement process with a maximum of 10 iterations over their remaining stages.

Table~\ref{tab:pass1_no_cost} reports the Pass@1 results of REA-Coder and its two variants on all 20 settings (4 LLMs $\times$ 5 benchmarks). REA-Coder consistently outperforms $REA\text{-}Coder_{WO-QA}$ in all settings. On average, REA-Coder improves over $REA\text{-}Coder_{WO-QA}$ by 5.82\%, with relative improvements ranging from 0.42\% to 12.57\%. This result shows that removing the question generation, answering, evaluation, and requirement revision steps leads to clear performance degradation. It suggests that these steps are effective in identifying misunderstood requirement details before code generation and revising the requirement accordingly.

REA-Coder also consistently outperforms $REA\text{-}Coder_{WO-MASK}$, and the performance drop is generally larger than that caused by removing the first group of steps. On average, REA-Coder improves over $REA\text{-}Coder_{WO-MASK}$ by 9.99\%. Across the five benchmarks, the average relative improvement ranges from 4.65\% to 17.14\%. This result shows that removing the masking, recovery, evaluation, and further refinement steps has a stronger negative impact on performance. These steps help check whether the generated code faithfully reflects the aligned requirement and provide useful signals for subsequent refinement.

\vspace{1mm}
\begin{custommdframed}
\textbf{Finding 2:} Both groups of steps make substantial contributions to the effectiveness of REA-Coder. In particular, removing the masking, recovery, evaluation, and refinement steps causes a larger performance drop than removing the question generation, answering, evaluation, and requirement revision steps.
\end{custommdframed}
\vspace{0mm}


\subsection{RQ3: Influence of the Number of Iterations}
To answer RQ3, we investigate how the number of iterations affects the effectiveness of REA-Coder. Since this RQ focuses on iterative refinement, we only include baselines that adopt an iterative fixing strategy, namely Self-Collaboration, Self-Repair, $\mu$Fix, and Specine. We vary the maximum number of iterations from 1 to 10 and evaluate the corresponding Pass@1 performance across different LLMs and benchmarks.

Figure~\ref{fig:iteration_ablation} shows that increasing the number of iterations consistently improves the performance of REA-Coder. Averaged over all 20 settings, REA-Coder achieves an average improvement of 19.80\% from iteration 1 to iteration 10. This result indicates that iterative requirement alignment is effective for improving code generation quality. In particular, we observe that even when the maximum number of iterations is set to only 1, REA-Coder already outperforms all compared baseline methods in terms of Pass@1 performance. This result fully demonstrates that the upfront requirement alignment process conducted prior to code generation can inherently and effectively boost the foundational performance of code generation.

We further observe that the performance gain is much larger in the early iterations, and then gradually becomes smaller. In particular, the average improvement from iteration 1 to 5 is 12.18\%, whereas the additional improvement from iteration 5 to 10 is only 6.29\%. This suggests that most requirement misunderstandings are corrected in the early iterations, while later iterations mainly provide limited refinement.

In addition, the performance of several baselines becomes much flatter in the later iterations, suggesting that their gains gradually saturate after the early rounds. By contrast, REA-Coder continues to achieve steady improvements as the number of iterations increases. This indicates that REA-Coder can better exploit additional iterations for further requirement alignment and refinement.

\vspace{1mm}
\begin{custommdframed}
\textbf{Finding 3:} Increasing the number of iterations consistently improves the effectiveness of REA-Coder. Compared with several baselines whose performance saturates in the later iterations, REA-Coder continues to improve steadily.
\end{custommdframed}
\vspace{0mm}
\vspace{1mm}

\begin{table*}[t]
\centering
\caption{The performance of zero-shot and REA-Coder\textsubscript{f} across LLMs and benchmarks.}
\resizebox{0.83\linewidth}{!}{
\small
\begin{tabular}{llccccc}
\toprule
\multirow{2}{*}{LLM} & \multirow{2}{*}{Method} & \multicolumn{5}{c}{Pass@1} \\
\cmidrule(lr){3-7}
& & APPS & CodeContests-raw & CodeContests & xCodeEval & LiveCodeBench-Lite \\
\midrule
\multirow{2}{*}{DeepSeek-v3.2}
& zero-shot & 66.00\% & 38.79\% & 37.20\% & 50.33\% & 40.57\% \\
& \textbf{$REA\text{-}Coder_f$} & \textbf{73.00\%} & \textbf{51.51\%} & \textbf{49.09\%} & \textbf{63.67\%} & \textbf{57.14\%} \\
\midrule
\multirow{2}{*}{Qwen3-Coder}
& zero-shot & 16.00\% & 16.97\% & 15.15\% & 9.67\% & 16.57\% \\
& \textbf{$REA\text{-}Coder_f$} & \textbf{49.67\%} & \textbf{23.03\%} & \textbf{18.18\%} & \textbf{43.00\%} & \textbf{35.43\%} \\
\midrule
\multirow{2}{*}{GPT-5-mini}
& zero-shot & 79.33\% & 61.82\% & 45.45\% & 54.67\% & 48.00\% \\
& \textbf{$REA\text{-}Coder_f$} & \textbf{81.00\%} & \textbf{65.45\%} & \textbf{56.97\%} & \textbf{64.00\%} & \textbf{57.71\%} \\
\midrule
\multirow{2}{*}{Gemini-3-Flash}
& zero-shot & 79.00\% & 55.76\% & 53.33\% & 69.67\% & 57.14\% \\
& \textbf{$REA\text{-}Coder_f$} & \textbf{84.33\%} & \textbf{69.06\%} & \textbf{67.27\%} & \textbf{80.67\%} & \textbf{73.14\%} \\
\bottomrule
\end{tabular}
}
\label{tab:pass1_no_cost_zero}
\end{table*}

%% file: 6-Analysis.tex
\section{Analysis}

\begin{figure*}[htbp]
  \centering 
  \includegraphics[width=0.88\linewidth]{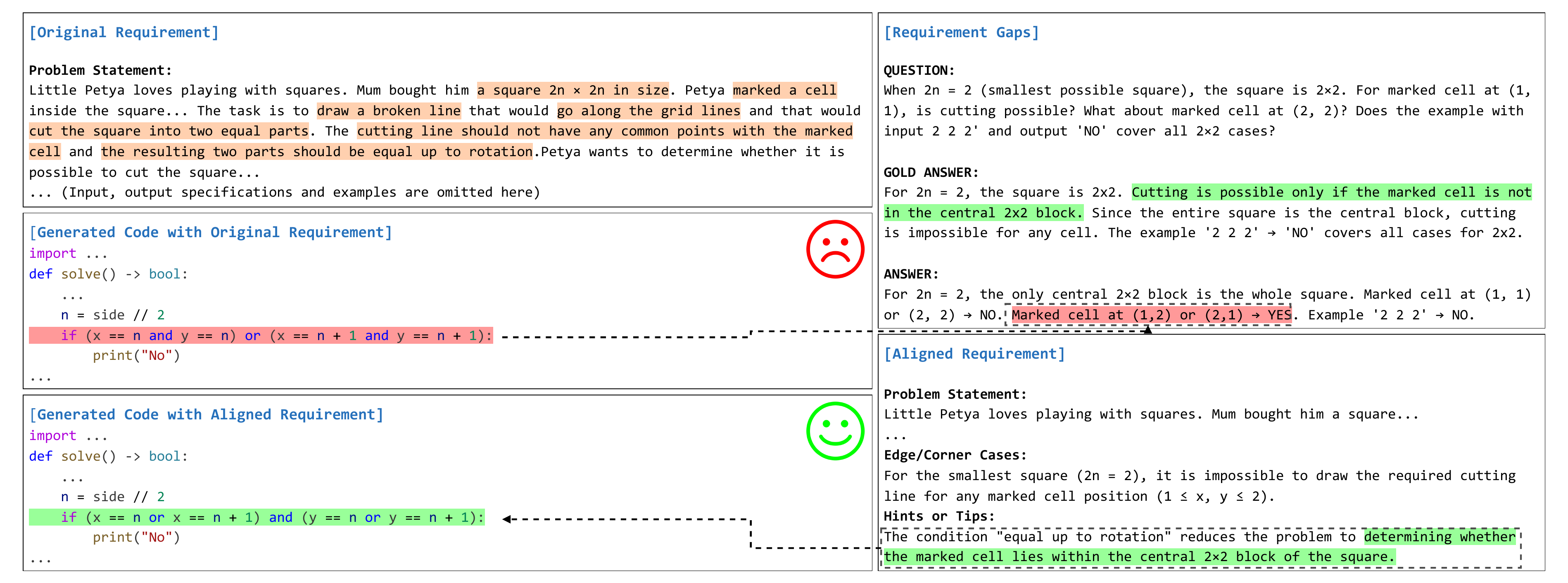}  
  \caption{Case Study of REA-Coder: An example of requirement alignment corrects an edge-case misunderstanding.} 
\label{fig:case_study_1}  
\end{figure*}

\begin{figure}[h]
  \centering 
  \includegraphics[width=\linewidth]{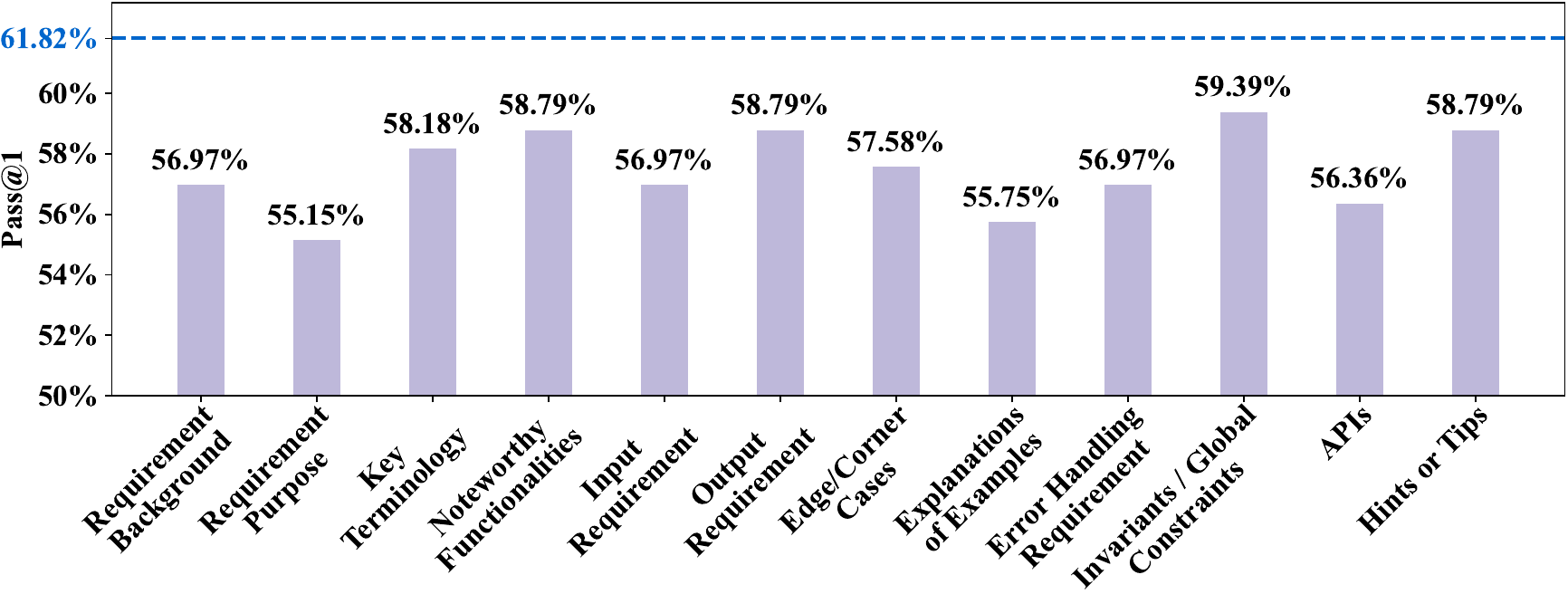} 
  \vspace{-6mm}
  \caption{Effectiveness of requirement dimension.} 
  \label{fig:parts_ablation}  
  \vspace{-4mm}
\end{figure}

\subsection{Performance of the First Code Generation} 
To examine whether REA-Coder can already improve requirement alignment at the initial generation stage and thereby produce more accurate code, we analyze the performance of its first code generation. We named this stage $REA\text{-}Coder_f$, which refers to the code generated after the requirement alignment based on question-answer-verification steps, without subsequent masking-based verification and iterative refinement.

Table~\ref{tab:pass1_no_cost_zero} reports the Pass@1 results of zero-shot and $REA\text{-}Coder_f$. Overall, $REA\text{-}Coder_f$ consistently outperforms zero-shot across all model and benchmark combinations in terms of Pass@1, showing that the benefit of our method emerges before multi-round refinement begins. 
On models with weaker code generation capability or datasets with more challenging tasks, the performance gains brought by REA-Coder are particularly substantial. For example, on Qwen3-Coder, REA-Coder improves first-round Pass@1 by 210.44\% on APPS, 113.82\% on LiveCodeBench-Lite, and 344.67\% on xCodeEval. On DeepSeek-v3.2, the improvements are also consistently strong, reaching 32.79\% on CodeContests-raw, 31.96\% on CodeContests, and 40.84\% on LiveCodeBench-Lite. On stronger models such as GPT-5-mini and Gemini-3-Flash, REA-Coder still delivers stable gains. For instance, it improves GPT-5-mini by 20.23\% on LiveCodeBench-Lite and 25.37\% on CodeContests, while improving Gemini-3-Flash by 25.01\% on LiveCodeBench-Lite and 23.91\% on CodeContests-raw.

The results validate the central design principle of REA-Coder: requirement alignment should be moved as early as possible in the code generation pipeline, rather than relying primarily on execution feedback after code has already been produced. Compared with methods that correct errors only after observing failed executions, REA-Coder enhances the requirement \emph{before} the code generation, allowing the model to generate code on an aligned requirement.

\subsection{Influence of Dimension}
To better understand which requirement dimensions contribute most to REA-Coder, we conduct an analysis. Specifically, for each run, we remove one dimension from the entire alignment pipeline, while keeping all other stages unchanged. We then measure the resulting overall performance and compare it with the full REA-Coder setting.

Figure~\ref{fig:parts_ablation} shows that removing any single dimension reduces performance. This consistent degradation indicates that all dimensions contribute positively to requirement alignment, and that the gains of REA-Coder do not come from a single dominant factor. Among all dimensions, Requirement Purpose has the largest impact. Removing it reduces performance to 55.15\%, a drop of 12.09\%. This suggests that explicitly clarifying what the program is expected to accomplish is central to correct code generation. The second largest drop is from removing Explanations of Examples, which lowers performance to 55.75\%. This highlights the importance of examples not merely as demonstrations of input and output pairs, but as concrete explanations of the underlying transformation logic. This finding suggests that developers should focus on these two aspects when specifying requirements for code generation. In the future, we plan to explore more effective alignment rules for better requirement alignment.

\subsection{Case Study}
To further understand why REA-Coder improves code generation, we conduct a qualitative case study focusing on the two core stages of our approach: requirement alignment and requirement alignment verification. We use two representative examples to illustrate how REA-Coder progressively identifies requirement misunderstandings, generates enhanced requirement, and finally improves the generated code. 

Figure~\ref{fig:case_study_1} presents a representative case where direct code generation from the original requirement produces an incorrect solution, while REA-Coder succeeds after the requirement alignment stage. The task asks whether a \(2n \times 2n\) square can be cut into two centrally symmetric parts without the cut touching the marked cell. The model does not correctly understand the centrally symmetric constraint, and thus misjudges the smallest \(2 \times 2\) case. This misalignment is reflected in the generated code, which uses an incorrect condition for the marked cell. 
REA-Coder identifies this error through question answering, clarifies the edge case and the equivalent condition in the enhanced requirement, and thereby guides the model to generate correct code.

\begin{figure*}[htbp]
  \centering 
  \includegraphics[width=0.88\linewidth]
  {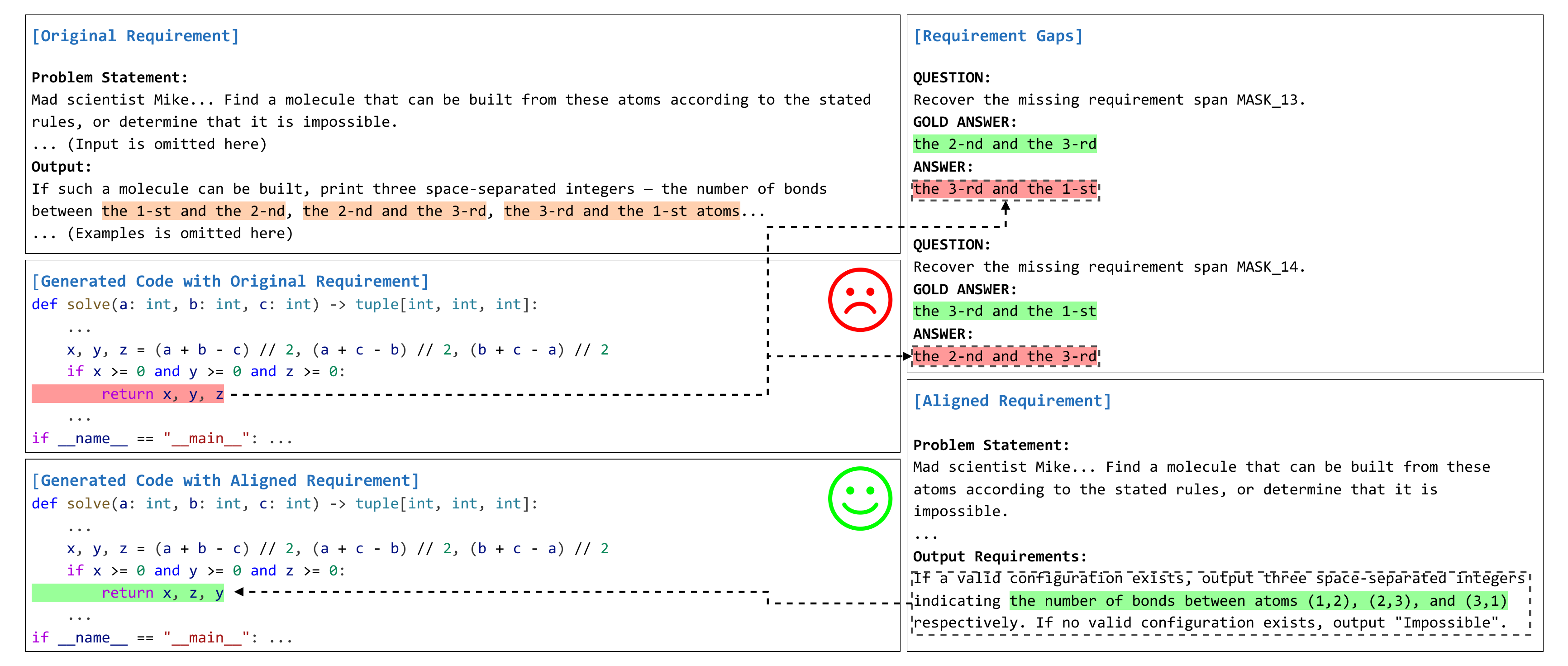}  
  \caption{Case Study of REA-Coder: An example of requirement alignment verification corrects output-order misalignment.} 
  \label{fig:case_study_2}  
\end{figure*}

Figure~\ref{fig:case_study_2} presents a second representative case where only requirement alignment stage is still insufficient. The task requires outputting the numbers of bonds between atom pairs \((1,2)\), \((2,3)\), and \((3,1)\) in this exact order. Although the model captures the main solution logic after the requirement alignment stage, it still fails to align with the output requirement and generates the last two numbers in the wrong order. REA-Coder exposes this remaining misalignment by masking some parts of the requirement and asking for recovering them from the generated code. Based on the recovery error, REA-Coder further strengthens the requirement in the next round, which finally leads to correct code generation.

\subsection{Threats and Validity}
To mitigate external threats, we evaluate REA-Coder on four different LLMs and five widely used code generation benchmarks, covering diverse problem characteristics and difficulty levels. In addition, we adopt Pass@1 as the main evaluation metric, which is widely used in code generation research and directly measures functional correctness through test execution. We also compare REA-Coder against a diverse set of strong baselines spanning reasoning-based methods and post-processing methods, demonstrating that our performance gains are consistent across different scenarios and not limited to a specific setting. 

The internal validity mainly lies in the used LLMs. Following common practice in recent studies, we mitigate this threat by carefully controlling the experimental setup, including the model configuration, prompting procedure, and iteration budget. In particular, we set the temperature to 0 for code generation, 0.2 for other steps, and 10 for the maximum iteration number. Moreover, we repeat each experiment three times and report the average results, which helps reduce the influence of random variation in model outputs. 
We conduct the significance test and the p-value of the experimental results is less than 0.05.
Furthermore, we only use public test cases to determine whether to stop iterations and leave out the execution feedback to LLMs, aiming to evaluate the effectiveness of requirement alignment that is the focus of this paper.

%% file: 2-Related_Work.tex
\section{Related Work}

Code generation aims to automatically generate executable programs from natural language requirements. 
Recent advances in LLMs have stimulated extensive research on improving code generation \cite{jiang2026survey, chen2021evaluating, austin2021program}. Existing methods can be broadly divided into two lines of work: reasoning-based generation and post-processing improvement.

\textbf{Reasoning-based generation} aims to improve code quality by enhancing the intermediate reasoning process before code is produced \cite{han2025multi, jiang2024self}. A representative direction is to design better prompting strategies for structured reasoning. For example, SCoT~\cite{li2025structured} introduces structured chain-of-thought prompting to guide the model through explicit intermediate steps prior to code generation. RoutingGen further proposes ICoT~\cite{li2025intention}, a difficulty-aware routing framework that adaptively selects prompting strategies according to task complexity and invokes reasoning for harder problems. Another direction is to organize code generation into multiple collaborative stages \cite{huang2312agentcoder, hong2023metagpt, qian2024chatdev}. Self-Collaboration~\cite{dong2024self}, for instance, decomposes the process into planning, coding, testing, and iterative revision, enabling the model to reason and refine in a more structured workflow \cite{mao2025blueprint2code}. These methods mainly focus on improving how the model reasons about the task and translates its reasoning results into code \cite{huang2023codecot}.

\textbf{Post-processing improvement} focuses on improving code generation after an initial code solution has been produced \cite{chen2023teaching, zheng2024opencodeinterpreter, wang-etal-2022-compilable}. One branch directly repairs generated code using execution feedback \cite{dou2024stepcoder, grubisic2024compiler, peng2025perfcodegen}. For example, Self-Edit~\cite{zhang2023self} converts execution-time fault information into targeted editing signals for code correction. Another branch further considers the mismatch between the generated result and the requirement, and attempts to refine either the code or the requirement accordingly \cite{zhou2025refinecoder, chen-etal-2025-revisit}. ClarifyGPT~\cite{mu2023clarifygpt} detects misaligned requirements through code consistency checks. $\mu$FiX~\cite{tian2025fixing} first outputs code based on the original requirement and mitigates specification misunderstanding by combining stronger reasoning prompts with feedback-driven correction. Specine~\cite{tian2025aligning} also previously generates code and iteratively supplements the requirement. SpecFix~\cite{jia2025automated} analyzes behavioral divergence among generated solutions and produces minimal textual patches to repair ambiguous problem descriptions. These methods show that refining the generated code or refining the requirement after generation can effectively improve performance \cite{le2022coderl}.

Despite their differences, these approaches share a common assumption: the model has already understood the given requirement correctly, and the remaining challenge is how to reason better or how to repair the generated result more effectively. However, this assumption does not always hold. In practice, LLMs may misunderstand the requirement itself before code generation begins. Under such requirement misalignment, even strong reasoning strategies or powerful post-processing mechanisms may still fail.  
This limitation motivates our work, which aligns requirements before code generation and verifies whether requirements have been understood through requirement alignment verification. 

%% file: 7-Conclusion.tex
\section{Conclusion}
In this paper, we propose REA-Coder, a requirement alignment approach to enhance code generation performance of LLMs. 
We evaluate REA-Coder on four LLMs and five benchmarks. Experimental results show that it consistently outperforms four state-of-the-art baselines. 
The results demonstrate that requirement alignment should be moved as early as possible, being consistent with the central design principle of REA-Coder.